\documentclass[aps,prb,preprint,groupedaddress]{revtex4-2}

\usepackage{color}
\usepackage{graphicx}
\usepackage{dcolumn}

\begin{document}


\title{A first-principles study of Zn induced liquid metal embrittlement at bcc and fcc grain boundaries}


\author{Ujjal Saikia$^1$}
\email{u.saikia@mpie.de}
\author{Mira Todorova$^1$}
\email{m.todorova@mpie.de}
\author{Tilmann Hickel$^{1,2}$}
\email{tilmann.hickel@bam.de}

\affiliation{$^1$Max-Planck-Institut f{\"u}r Eisenforschung GmbH, Max-Planck-Str.~1, D-40237 D{\"u}sseldorf, Germany}
\affiliation{$^2$Bundesanstalt f{\"u}r Materialforschung und -pr{\"u}fung (BAM), Richard-Willst{\"a}tter-Str.~11, 12489 Berlin, Germany}



\begin{abstract}
Zn induced liquid metal embrittlement (LME) is a major concern in particular for advanced high strength steels, which often contain a significant amount of austenite compared to established steel grades. Using density functional theory (DFT) calculations we, therefore, compare the behaviour of Zn in ferrite (bcc) and austenite (fcc) grain boundaries (GBs) with different magnetic ordering to investigate the role of crystal structure as well as magnetism in LME. We address the performance of DFT based paramagnetic calculations by utilizing the spin space averaging relaxation approach. Our results show that both magnetic and elastic contributions have significant influence towards segregation and embrittling behaviour of Zn. The primary requirement is the elastic contribution, while the presence of magnetic disorder increases the critical concentrations for the onset of GB weakening. While Zn segregation is more favourable in bcc compared to fcc GB, larger impact of Zn coverage on GB weakening is observed for fcc.  For both structures, the rapid decrease in surface defect state energies is identified as the driving force behind GB weakening. These surface defect states stabilize at lower Zn concentrations than GB defect states.
\end{abstract}


\maketitle

\section{Introduction}
The development of advanced high strength steels (AHSS) is of particular interest for the automotive industry, because they can be used to reduce vehicle weight without compromising passenger safety. Within this strategy, medium and high Mn steels constitute an important materials system that shows promise for applications by virtue of improved balance of strength and elongation and has already given rise to many interesting materials discoveries. One of them is related to the fact that steel body parts of vehicles are commonly coated with Zn for corrosion protection. It has been frequently observed that the welding (e.g., resistance spot welding) of zinc-coated high-strength steels can lead to crack formation in the heat affected zone (see \cite{bhattacharya2018liquid} and references therein). The current assumption is that the decohesion of grain boundaries (GB) is caused by the formation of a liquid film of a low-melting metal, which is Zn in this case, resulting in a brittle transition of the material. The reduced elongation to failure ratio of an otherwise ductile metal due to the local formation of liquid metals is also known as liquid metal embrittlement (LME). 

The most important design strategy of AHSS with a large Mn content is the substantial increase of the amount of (retained) austenite as compared to conventional steels, through which formability as well as crash performance are enhanced by transformation (TRIP) or twinning (TWIP) induced plasticity. The increased amount of austenite could be a central reason why LME is a severe phenomenon for some AHSS, but this has not yet been confirmed unambiguously. Generally, it was reported that the sensitivity of steels to LME increases with their strength \cite{bhattacharya2018liquid, kang2016zn}. However, it is not clear, if this trend is driven by the intrinsic high strength of the material, which facilitates high stress concentrations at grain and phase boundaries prior to plastic deformation, the alloying designs needed to achieve high strength, or by the crystalline phase constituents and their interfaces \cite{bhattacharya2018liquid}. 
Specifically, C.~Beal \textit{et al.} \cite{beal2011liquid,beal2012embrittlement} observed that TWIP steels are more sensitive to LME than ferritic steels, whereas H.~Kang \textit{et al.} \cite{kang2016zn} observed no dependence of Zn-induced LME on the austenite fraction in high strength steels. Hence, it is a central goal of this study to determine whether the austenitic phase gives rise to a higher sensitivity to Zn-induced LME in AHSS as compared to the ferritic phase.

Furthermore, it is commonly believed that LME is caused by a change in the electronic structure or the local atomic arrangement due to the presence of solute atoms at a GB \cite{luzzi1991atomic,yan1993interatomic,heo2013grain,cueto2012macroscopic}. Based on first-principles data, Gibson \textit{et al.} \cite{gibson2016survey} showed that GB cohesion depends on the bond energy formed by a solute. Lejcek \textit{et al.} \cite{lejvcek2017interfacial} performed an extensive quality assessment of the experimental and theoretical results on solute segregation in bcc-Fe and Ni. A substantial amount of atomistic simulation studies on the interface structure and their impact on the microstructure evolution, GB premelting and GB diffusion is available in the literature \cite{mishin2010atomistic,schmidt_prediction_1995,tschopp_structure_2007,wu_grain_2009,suzuki_atomic_2005,williams_thermodynamics_2009,sutton_interfaces_1995,broughton_thermodynamic_1986,williams_thermodynamics_2009-1,schonfelder_comparative_2005}. Two of the suggested mechanisms are the formation of strong directional bonds that are reducing dislocation motion during decohesion and structural-volumetric effects. However, most of these studies were performed with empirical interatomic potentials and hence, lack the advantages of first-principles approaches to adequately treat the electronic degrees of freedom and to have a high predictive power.

To get deeper insight of the LME mechanisms without these shortcomings, K.~Bauer \textit{et al.} \cite{bauer2015first} performed a first-principles study on Zn interaction with two tilt GBs in bcc Fe. This study nicely demonstrates the insights that can be achieved by first-principles methods in this context. The strong preference of the Zn atoms for open surfaces over GBs was predicted as the main thermodynamic driving force for decohesion. Since kinetic barriers in Zn diffusion, which depend on the surface orientation and termination, could reduce the undesired effect of GB weakening, their consideration is essential for obtaining a complete understanding of the LME process. 
D. ~Scheiber \textit{et al.} \cite{scheiber2020influence} investigated the segregation of Zn and two other alloying elements (Al and Si) in three bcc Fe GBs. They reported a stronger tendency of Zn to enrich the GBs and surfaces than Al and Si. A strong negative effect of Zn, compared to a small negative effect of Al and a small positive effect of Si towards GB cohesion was also predicted in their study. 

Only a few ab initio studies are available on solute segregation to paramagnetic (PM) grain boundaries owing to the difficulty of handling PM within the framework of density functional theory (DFT). Pure Fe has a bcc crystal structure and is ferromagnetically (FM) ordered at room temperature, but shows already at 100 $^\circ$C a reduction of the magnetization by about 10 \% before it eventually transforms into the PM bcc phase at 770 $^\circ$C. At 910 $^\circ$C this phase transforms into the PM fcc crystal structure. If Fe is constrained to the fcc structure, it prefers the PM disorder also well below room temperature. 
These structural and magnetic phase transitions and the stabilities of phases are significantly influenced by different alloying elements used in the steel. 
For example, Mn is not only known to reduce the ferrite-austenite transformation temperature, but also the Curie temperature of magnetic ordering \cite{schneider_ab_2021, paduani_mossbauer_2014, arajs_ferromagnetic_1965, li_characteristics_2002, bigdeli_thermodynamic_2019}. In the suitable temperature range for Zn induced LME, the most relevant magnetic configuration for Fe-Zn would be paramagnetic disorder. Therefore, in addition to an understanding of the segregation behaviour of Zn in FM Fe, it is crucial to investigate the same also in PM Fe.

The previous works do not address the behavior of Zn in fcc Fe GBs, which is highly relevant in the context of LME in AHSS, as temperatures can be well above Ac$\rm{_1}$ in the crack sensitive regions of spot welds in the heat affected zone. Another important piece missing from the picture is the effect of the magnetic configuration on the Zn behavior. Our previous experience with Mn segregation in paramagnetic bcc GB gives us a strong reason to believe that magnetism could be an important factor in determining the segregation behaviour of Zn in Fe GBs. Therefore, prime objective of this work is to systematically investigate the role of ferrite (bcc) and austenite (fcc) phases of Fe in determining the behaviour of Zn towards LME in AHSS with significant austenite content. We considered both FM and PM configurations for bcc GB, and non-magnetic (NM) and PM for fcc GB. In this study, we employ our recently developed spin-space averaging (SSA) relaxation technique \cite{hegde2020atomic, kormann2012atomic} to treat the PM grain boundaries.

\section{Computational methods}
The first principles calculations are performed with density functional theory (DFT) \cite{hohenberg1964inhomogeneous} as implemented in the Vienna Ab-initio Simulation Package (VASP) \cite{kresse1996efficient}. We use projector-augmented wave (PAW) potentials \cite{perdew1992accurate} with the generalized gradient approximation of Perdew-Burke-Ernzerhof (PBE) \cite{perdew1996generalized}. A kinetic energy cutoff of 400 eV for plane waves with a $k$-mesh equivalent to (18$\times$18$\times$18) for bulk Fe and (28$\times$28$\times$28) for bulk Zn are used. Ionic relaxations are performed until the magnitudes of the force components on all atoms are less than 0.01 eV/{\AA}. With these parameters the GB and surface energies are converged to within 2 meV/{\AA}$\rm{^2}$. The optimized bulk Fe lattice parameters used in this study are 2.832 {\AA}, 2.831 {\AA}, 3.446 {\AA} and 3.50 {\AA} for bcc FM, bcc PM, fcc NM and fcc PM respectively. We have used a super-cell approach with periodically repeated orthorhombic cells to construct the GBs and surfaces.

We calculate the GB energies and corresponding surface energies as 
\begin{equation}\label{energy_gb}
\gamma_\mathrm{GB} = (E_{N_\mathrm{Zn}@\mathrm{FeGB}} - N_\mathrm{Fe}\mu_\mathrm{Fe} - N_\mathrm{Zn}\mu_\mathrm{Zn})/2A_\mathrm{GB}
\end{equation}
\begin{equation}\label{energy_surf}
\gamma_\mathrm{Surf} = (E_{N_\mathrm{Zn}@\mathrm{FeSurf}} - N_\mathrm{Fe}\mu_\mathrm{Fe} - N_\mathrm{Zn}\mu_\mathrm{Zn})/2A_\mathrm{Surf}
\end{equation}

Here, $E_{N_\mathrm{Zn}@\mathrm{FeGB}}$ is the total energy of the Fe GB with $N_\mathrm{Zn}$ number of iron atoms exchanged between the GB and a Zn reservoir, characterized by its chemical potential $\mu_\mathrm{Zn}$. The term $N_\mathrm{Fe}$ represents the number of iron atoms exchanged between the GB and the iron reservoir, characterized by its chemical potential $\mu_{Fe}$. The chemical potentials, i.e. reservoirs, for the iron and zinc atom are taken to be the respective bulk phases of these materials. Similarly, $E_{N_\mathrm{Zn}@\mathrm{FeSurf}}$ is the total energy of the Fe surface slab with $N_\mathrm{Zn}$ iron atoms substituted by zinc. The factor $1/2$ accounts for the presence of two identical grain boundaries or surfaces within our simulation cell, a consequence of the periodic boundary conditions we use. Here, $A_\mathrm{GB}$ and $A_\mathrm{Surf}$ correspond to the area of one GB plane, and respective surface.

The Zn induced embrittling behaviour was investigated within the Rice-Wang model as
\begin{equation}\label{rice_wang}
\Delta\gamma = 2\gamma_\mathrm{Surf}(\mu_\mathrm{Zn}) - \gamma_\mathrm{GB}(\mu_\mathrm{Zn}).
\end{equation}

The segregation energy of a single Zn atom to the GB is given by
\begin{equation}\label{seg_first_zn}
E_\mathrm{seg} = [E_\mathrm{Zn1@FeGB} - E_\mathrm{FeGB}] - [E_\mathrm{Zn1@FeBulk} - E_\mathrm{FeBulk}].
\end{equation}
Here, $E_\mathrm{Zn1@FeGB}$ is the total energy of the GB with one Fe atom substituted by Zn, while the total energy of the pristine GB and bulk systems are denoted by $E_\mathrm{FeGB}$ and $E_\mathrm{FeBulk}$ respectively.

The segregation energy of a second Zn atom to the GB when one Zn is already present in the GB is given by
\begin{equation}\label{seg_second_zn}
E_\mathrm{seg} = [E_\mathrm{Zn2@FeGB} - E_\mathrm{Zn1@FeGB}] - [E_\mathrm{Zn1@FeBulk} - E_\mathrm{FeBulk}]
\end{equation}
Here, $E_\mathrm{Zn2@FeGB}$ is the total energy of the GB in which  two Fe atoms are substituted by Zn.

There are several methods to describe PM in ab initio calculations, each of which has its advantages and disadvantages \cite{Fritz_Lambda_251}. Here, we follow the common strategy of using a collinear approximation, where the distribution of up- and down-spin on the atoms is defined by special quasi-random structures (SQSs). This approach has been found to yield reliable results for bulk systems without any defects. However, in systems containing defects, such as GBs, a single SQS structure typically does not provide sufficient statistical representation of the spin configurations at the defect. The pragmatic approach to use an antiferromagnetic double-layer (AFMD) to approximate the PM configuration of Fe, as occasionally done in the literature \cite{boukhvalov2007magnetism,medvedeva2010magnetism}, is an even more severe restriction. In this study we adopted, therefore, the spin-space averaging (SSA) technique originally developed for phonon calculations of perfect bulk structures to describe the PM configurations of GBs \cite{kormann2012atomic}. This ensures, that the energy and the relaxation of the Fe atoms adjacent to the GB is based on the DFT results of several SQS spin configurations. The core approximation of this approach is that magnetic configurations in the PM state fluctuate so quickly that an atom does not have sufficient time to respond to any single magnetic configuration. Hence, instead of an instantaneous force resulting from a static magnetic configuration the atoms are displaced according to the averaged force over several (here 15) different magnetic configurations. We use electronic structure code SPHInX \cite{boeck_object-oriented_2011} for DFT calculation of different SQS spin configurations which allows us to employ an efficient method for spin constraints \cite{hegde2020atomic}. The “external structure optimizer” (SxExtOpt) \cite{freysoldt_fly_2017} is used for on the fly structural relaxations after each SSA step. In this approach, the atoms undergo relaxation based on spin-averaged forces, ensuring the preservation of symmetry. This guarantees a relaxation within the adiabatic limit, where spins fluctuate rapidly and atoms move slowly. Consequently, atomic positions are adjusted until convergence of SSA forces is reached. Further details about the SSA relaxation method is available in \cite{hegde2020atomic}. 
We have used the Python-based integrated development environment (IDE) pyiron \cite{pyiron-paper}, leveraging its capabilities to seamlessly execute complex simulation protocols.

\section{Results}
\subsection{Influence of Zn on the grain boundary structure}

\begin{figure}
\centering
\includegraphics[width=0.5\textwidth]{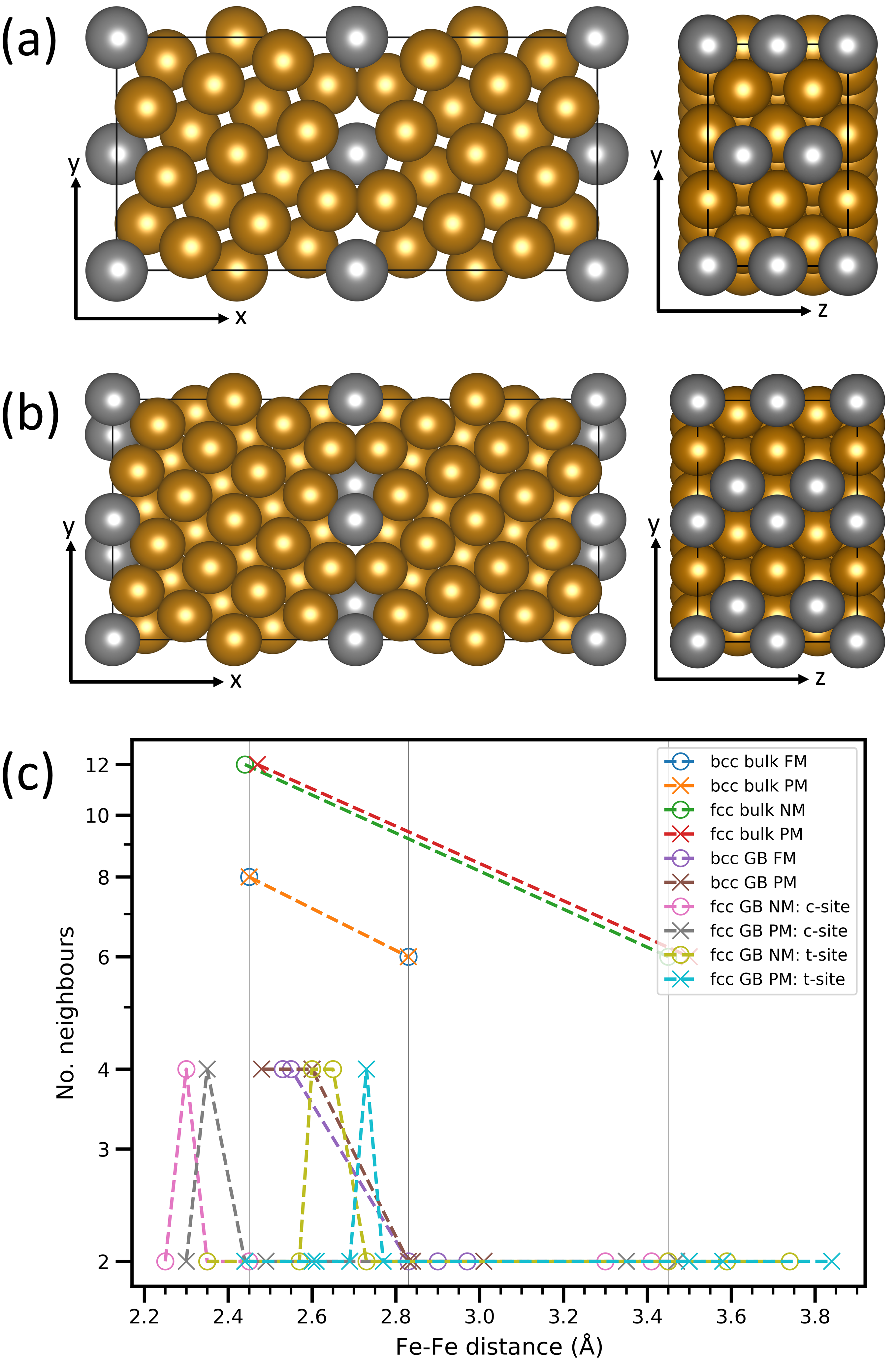}
\caption{\label{structure} The structure of the (a) bcc and (b) fcc $\Sigma 5[100]36.8^{\circ}$ symmetric tilt grain boundaries used for simulations. The atoms in the GB plane are represented by gray coloured spheres. (c) The distribution of nearest neighbours for a Fe atom sitting in the GB plane. For reference, the distance of nearest and second-nearest neighbor atoms is also depicted for a Fe atom sitting in a bulk bcc and a bulk fcc environment. By construction there are four equivalent sites in the bcc GB plane and overall eight sites in the fcc GB plane belonging either two of the inequivalent sites, i.e., a compression site (c-site) or a tensile site (t-site).}
\end{figure}

We have used the coincidence site lattice approach \cite{humphreys2012recrystallization} to construct the bcc and fcc $\Sigma 5[100]36.8^{\circ}$ symmetric tilt grain boundaries (STGBs) shown in Fig.~\ref{structure} (a) and (b). By construction, the grains are periodic along the $yz$-plane with the corresponding bulk-stacking order lattice planes tilted with respect to the $x$ direction of a conventional unit cell. This stacking order is reversed after half a period in $x$ direction, forming a tilt boundary between the two grains. The chosen GB has a more open geometry compared to bulk-like GBs such as $\Sigma 3$ and is discussed in the experimental literature as a boundary along which Fe-alloys tend to embrittle \cite{noskovich_alloy_nodate}.

The $\Sigma 5$ GB structure is constituted by two \{013\} planes. In our simulation cell for the FM and PM bcc GB, the interface plane has an area of 50.72 {\AA}$^{2}$ and contains four atoms per layer perpendicular to the $x$ direction. The atomic area density of this plane is 37\% smaller (0.08 Atoms/{\AA}$^{2}$) than that of a bcc (100) plane. The Fe atoms sitting at the GB plane have a 13\% higher Voronoi volume compared to bulk bcc Fe. 

The fcc GB interface plane of the supercell shown in Fig.~\ref{structure}(b) contains eight atoms per layer perpendicular to the  $x$ direction and has an area of 75.10 {\AA}$^{2}$ for the NM and 77.48 {\AA}$^{2}$ for the PM state. Its atomic area density is 58\% smaller (0.11 atoms/{\AA}$^{2}$) than that of the fcc (100) plane for both, the NM and PM configuration. 
The Fe atoms at the fcc $\Sigma 5$ GB plane occupy two inequivalent sites, a compression site (c-site: site 0a in Fig. \ref{co_segregation} (b)) and a tensile site (t-site: site 0b in Fig. \ref{co_segregation} (b)). For the  fcc NM GB, the t-site has a 17\% larger and c-site has a 7\% smaller Voronoi volume compared to bulk fcc NM Fe. For the fcc PM GB, the t-site has a 19\% larger and c-site has a 4\% smaller Voronoi volume compared to bulk fcc PM Fe.

The nearest neighbour (NN) environment of the atoms at the GB plane is also very different compared to their bulk counterparts. As shown in Fig.~\ref{structure} (c), Fe atoms in their bulk bcc environment have eight first NN at 2.45 {\AA} and six second NN at 2.83 {\AA}. For a Fe atom at the bcc FM (PM) GB plane, eight first NN atoms are distributed within a distance range 2.53-2.55 {\AA} (2.48-2.60 {\AA}) and six second NN atoms are distributed between 2.83-2.97 {\AA } (2.83-3.01 {\AA}). For fcc NM (PM), a bulk Fe atom has 12 first NN at 2.44 {\AA} (2.47 {\AA}) and six second NN at 3.45 {\AA} (3.50 {\AA}). The Fe atoms sitting at the c-site of the NM (PM) fcc GB have 10 first NN atoms distributed between 2.25-2.45 {\AA} (2.30-2.49 {\AA}) and six second NN distributed from 3.30-3.45 {\AA} (3.35-3.50 {\AA}). Whereas, for a t-site of NM (PM) GB 14 first NN atoms are distributed between 2.35-2.73 {\AA} (2.44-2.77 {\AA}) and six second NN are distributed between 3.45-3.74 {\AA} (3.50-3.84 {\AA}).

\begin{widetext}
\begin{figure}
\centering
\includegraphics[width=0.7\textwidth]{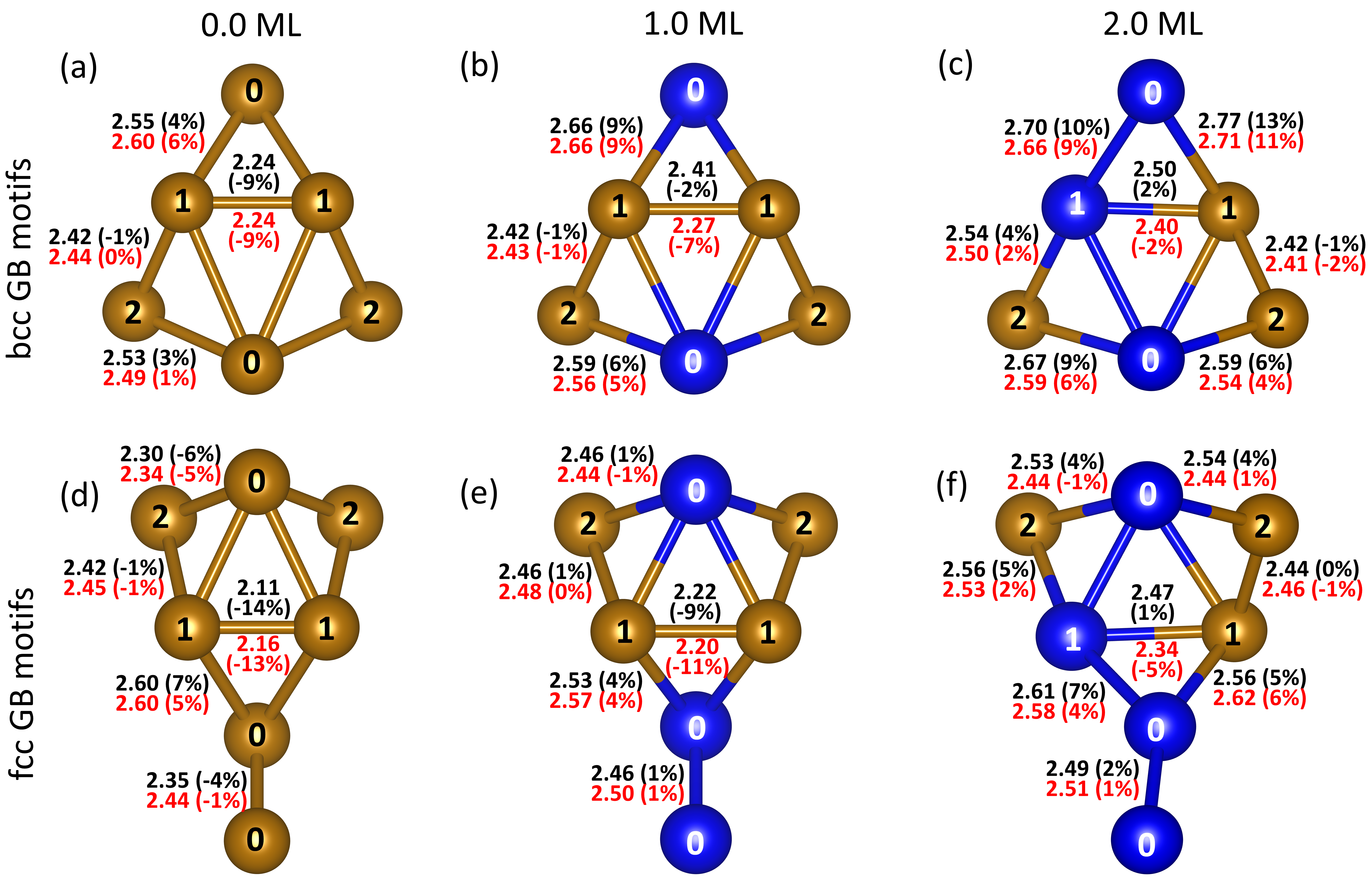}
\caption{\label{bond_length} The structure of a $\Sigma 5$ GB consists of kite-shaped motifs for bcc (top row) and kite-shaped motifs with a tail for fcc (bottom row). The atoms marked 0, 1, and 2 are sitting at the GB plane, the first neighbouring plane and the second neighbouring plane, respectively. The bond length for different Zn coverages (blue atoms) are given next to each bond. The values given in parenthesis are the relative change in bond length compared to a bulk Fe-Fe bond. A positive value implies an increase and a negative value implies a decrease in bond lengths. The bond length values for bcc FM and fcc NM are given in black colour. The red coloured values are for PM GBs. ML refers to monolayers of Zn.}
\end{figure}
\end{widetext}

The structure of the $\Sigma 5$ GB is composed of a distinct, reoccurring arrangement of kite-shaped motifs for bcc (Fig. \ref{bond_length} (a)) and kite-shaped motifs with a tail for fcc (Fig. \ref{bond_length} (d)). The Fe-Fe bond length in these motifs reflects the change in bond length in the GB environment compared to the corresponding bulk lattice. For pristine GBs, the most pronounced change was observed between the Fe atoms in site ``1", which is reduced by 9\% for both FM and PM bcc, and by 14\% (13\%) for fcc NM (PM). To understand the impact Zn has on the structure of the GB we incorporated Zn in both bcc and fcc GBs. Given the size similarities between Fe and Zn, one can expect the substitution of Fe atoms by Zn to be a more favourable scenario than Zn as interstitial inside the Fe matrix, as observed already by K.~Bauer \textit{et al.} \cite{bauer2015first} for the bcc Fe $\Sigma 5$ GB. 

The analysis of the geometric structure of the most stable Zn configuration in the $\Sigma 5$ GB, shown in Figs.~\ref{bond_length}(a)-(c) for bcc and in Figs.~\ref{bond_length}(d)-(f) for fcc, reveals fundamental differences between the Fe sites in the bcc and the fcc GB. The differences in Voronoi volume and NN density for Fe sites in the bcc and fcc GB were discussed in the preceding and suggest that one can expect differences in the segregation behaviour of Zn. Comparing the pristine GB with the respective GB covered by 1 or 2 monolayers (ML) of Zn, we observe that the Zn incorporation has a much  higher impact on the bond length' within the GB, compared to the impact caused by the change of the magnetic configuration. For both GBs we find that incorporation of Zn results in an elongation, i.e., softening of the bonds of atoms in the vicinity of the GB. Furthermore, bonds within the kite (e.g. (1,1)) are more contracted in the PM regime, as compared to the FM bcc and NM fcc case.  However, the relaxation behaviour within the fcc GB is stronger than in the bcc GB, which is very likely a consequence of the higher Voronoi volume and lower NN density of the Fe atoms within the bcc GB (compared to bulk).

\subsection{Segregation of Zn to the grain boundaries}

\begin{widetext}
\begin{figure}
\centering
\includegraphics[width=\textwidth]{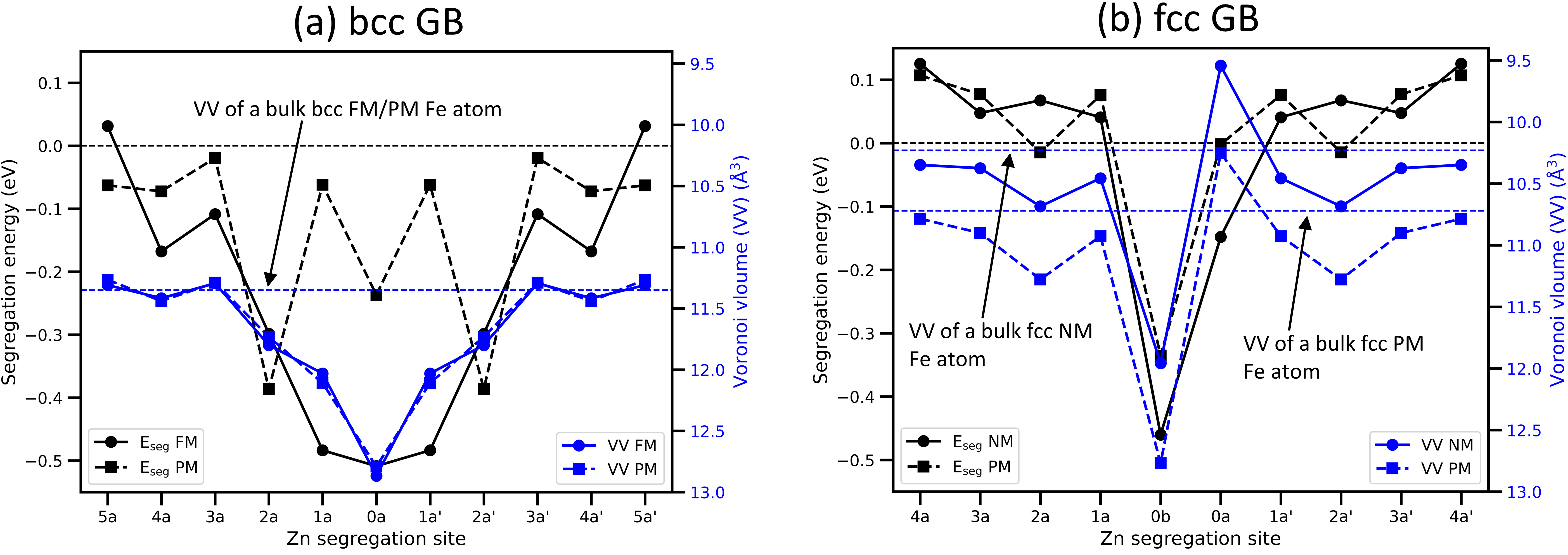}
\caption{\label{segregation}  Zn segregation energy ($\rm{E_{seg}}$) to different Fe sites across the GB plane (black curves) and Voronoi volume (VV) (blue curves) of those sites for (a) bcc FM and PM and, (b) fcc NM and PM $\Sigma 5$ GBs are depicted here. The numbers 0, 1, etc., in the $x$ tick labels indicate that the segregation sites is at the GB plane, next to the GB plane, etc., respectively. Please see Fig.~\ref{co_segregation} (c) and (d) for the exact location of the site labels in the bcc and fcc GB supercells, respectively.}
\end{figure}
\end{widetext}

In the context of LME, there are two possible sources for the presence of Zn at GBs: (i) Zn atoms from the coating that were already dissolved in iron bulk or (ii) Zn atoms that originate from the coating and penetrate the materials via GBs. In this section we focus on the first source, i.e., Zn atoms originating from the bulk of the material, since it is important to understand the impact of Zn segregation on GB prior to the LME induced crack initiation and propagation. The second source, i.e. Zn atoms originating from the coating,  and its impact on GB/Surface defect states energetics are the subject of the following section.

We have used Eq.~\ref{seg_first_zn} to calculate the segregation energy of a Zn atom from bulk Fe to the GB (Fig.~\ref{segregation}). A negative segregation energy value implies that moving a solute Zn atom from the bulk Fe matrix to the GB is connected with a gain in energy. As shown in Fig.~\ref{segregation} (a) for bcc FM GB (solid black curve), the most favourable segregation site is found in the GB plane (-0.5 eV), but even up to the $\rm{4^{th}}$ atomic layer from the GB plane the segregation is favourable. This indicates that Zn has a strong tendency of segregation to the bcc FM GB. Changing the magnetic configuration of the bcc GB from FM to PM leads to a significant reduction in segregation tendency for all considered sites. Exceptions are the ``2a" or ``2a$^{'}$" sites (black dashed curve in Fig.~\ref{segregation} (a)), where the segregation of Zn triggers a significant reconstruction of the GB structure which results in a higher segregation tendency compared to the other sites.

For fcc NM GB, Zn segregation is favourable only in the GB plane (black solid curve in Fig.~\ref{segregation} (b)). Between the two inequivalent sites in this GB plane, the site with the larger Voronoi volume (``0b'') is more favourable for Zn segregation (-0.45 eV) compared to the ``0a'' site (-0.15 eV), which has a smaller Voronoi volume compared to bulk fcc Fe. Changing the magnetic configuration from NM to PM leads to a decrease in segregation tendency for both sites with the Zn segregation energy in ``0b" becoming -0.34 eV and in ``0a" 0 eV (black dashed curve in Fig.~\ref{segregation} (b)). 

The impact of the magnetic state of Fe and of elastic contributions on the segregation behaviour of Zn is summarised in Fig.~\ref{segregation}. Considering the size similarity between Fe and Zn, if a Fe-site at the GB has higher Voronoi volume than a Fe-site in it's respective bulk phase, we consider that site at the GB to be elastically favourable for Zn segregation. Although, changing the magnetic configuration of the GB to PM results no change in Voronoi volume of the Fe atoms in bcc (compared to FM GB) and slight increase in Voronoi volume in fcc (compared to NM GB), we noticed a reduction of the Zn segregation tendency for both bcc and fcc. The effect is larger in a bcc GB compared to fcc GB, as discussed previously. The impact of the elastic contributions on the Zn segregation behaviour is intimately connected to the Voronoi volume. We calculated the Voronoi volume of the segregation sites, depicted in Fig.~\ref{segregation} (blue curves), using the respective pristine GB supercells and find for both bcc and fcc GBs a strong correlation between Voronoi volume and segregation energy. Specifically, for the site with the largest Voronoi volume, the Fe-site in the FM GB plane (blue solid curve in Fig.~\ref{segregation} (a)), which is 13\% larger than bulk bcc Fe-site, segregation is most favourable. Further away from the GB plane the Voronoi volume decreases gradually and segregation become less and less favourable. It is worth to note, that while the Voronoi volume of the Fe-sites in both the bcc PM and FM GBs is the same, the segregation energy changes significantly, implying that the difference in Zn segregation behaviour in the bcc PM GB compared to FM GB is directly connected to the magnetic interaction strength due to the magnetic state of Fe.

For fcc NM GB we find again that the site with the largest Voronoi volume (``0b'', 17\% larger than a bulk fcc Fe and 7\% smaller than a Fe in the bcc GB plane) is most favourable for segregation. Interestingly, the site ``0a" which has a 7\% smaller Voronoi volume than a bulk fcc Fe is still slightly favourable for segregation. To understand this behaviour, we have analysed the nearest neighbours distances of site ``0a'' and compared it with a bulk fcc Fe site as shown in Fig.~\ref{structure} (c). A bulk fcc Fe-site has 12 nearest neighbours at a distance of 2.44 {\AA}. The site ``0a'' has only 10 nearest neighbours distributed between 2.25-2.45 {\AA}.
Hence, to replace a Fe atom by Zn in fcc bulk, we have to disrupt 12 relatively strong Fe-Fe bonds and replace them with weaker Fe-Zn bonds \cite{bauer2015first}. The smaller number of nearest neighbours in the GB site ``0a" means, that only 10 Fe-Fe bonds needs to be disrupted, hence segregation of Zn is favourable for site ``0a". 
Considering the bulk lattice parameter for fcc PM, which is with 3.50 {\AA} slightly higher than for NM fcc (3.45 {\AA}), it is not astonishing the fcc PM GB sites have larger Voronoi volume compared to the NM GB sites. Hence, using the elastic contribution argument for the fcc PM GB, we would expect a higher segregation tendency than in NM GB. However, we have seen that the segregation tendency is decreased for PM GB, which indicates that similar to the bcc PM GB case, the reduction of the Zn segregation tendency in fcc PM GB is solely due to the change in the magnetic state of Fe.

The preceding discussion shows that both elastic and magnetic contributions are relevant in determining the Zn segregation tendency. To better understand how these contributions affect the Zn segregation tendency we consider the competition between them.
When we construct a GB, some atomic sites acquire larger volume compared to their voulme in the bulk. At this point, we should also keep in mind that the structural and magnetic effects are coupled with each other. For a Fe atom, bcc PM and FM states has same volume and a Fe atom in fcc PM state has larger volume than in NM. 
In our GBs, only those atomic sites which are favourable elastically are also always favourable sites for Zn segregation. Now, when we switch the magnetic configuration of the GBs to a disordered state (PM), the volume of the Fe atoms remain same for bcc and increased slightly for fcc. However, Zn segregation tendency decreased for both bcc and fcc. This indicates that disordered magnetic state (PM) will always have negative impact on Zn segregation even if a site is initially elastically favourable and we made it elastically more favourable by switching the magnetic configuration to PM state.
In other words, if a site is elastically not favourable for segregation, we can't  make it favourable by just changing the magnetic configuration of the system.

\begin{widetext}
\begin{figure}
\centering
\includegraphics[width=\textwidth]{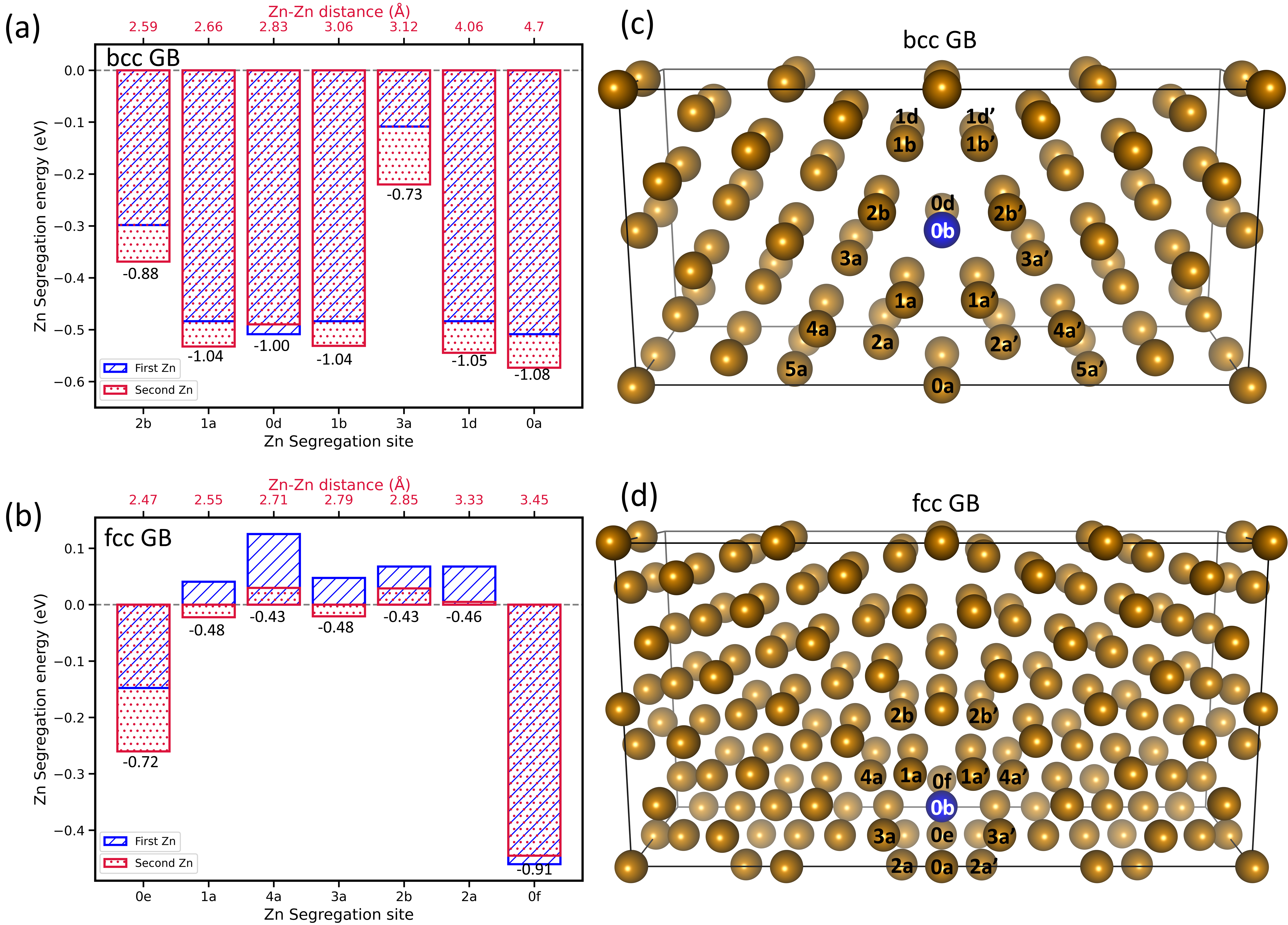}
\caption{\label{co_segregation} The segregation energy of first (blue bar) and second (red bar) Zn to (a) bcc FM and (b) fcc NM $\Sigma 5$ GBs. When we calculate the segregation energy of the second Zn, the first Zn is always sitting at its most favourable positions (blue spheres in (c) and (d)) which is site ``0b" in the both cases (for bcc GB, site ``0a" and ``0b" are equivalent). The values next to each bars indicate the total segregation energy of two Zn atoms. We insert the second Zn atom at different neighbouring sites of the first Zn as indicate by the lower $x$ axis. The site labels are marked in (c) for bcc and (d) for fcc GBs.}
\end{figure}
\end{widetext}

So far we have investigated the segregation of a single Zn to the pristine Fe GBs. In the context of LME it is important to understand the impact of coverage on the segregation behaviour of Zn to the GB. Since we have already learned, that elastic contributions are the primary requirement in determining the segregation of Zn in these GBs, to avoid computationally intensive PM calculations we study the co-segregation behaviour of Zn considering only the bcc FM and fcc NM GBs. Nevertheless, the knowledge obtained from studying the segregation of a single Zn atom to the bcc and fcc PM GBs allows us to infer that the co-segregation tendency we report in Fig.~\ref{co_segregation} will be slightly decreased for the respective PM GBs.

Assuming that a first Zn atom resides in the  most favourable (``0b'') site for both bcc and fcc GBs (blue spheres in Fig.~\ref{co_segregation} (c) and (d)), when adding the second Zn atom we consider all neighbouring inequivalent Fe sites around the first Zn up to the $\rm{3^{rd}}$ atomic layer from the GB plane. The segregation energy we calculate using Eq.~\ref{seg_second_zn} for a second Zn atom added to the system subsequently in each of the considered sites is plotted in Fig.~\ref{co_segregation} (red bars). For comparison, the segregation energy of the first Zn atom in each of these sites are also shown as blue bars. For the bcc GB, the presence of first Zn at site ``0b" favours segregation of a second Zn atom to all considered sites with the exception of site ``0d". The total segregation energy for two Zn (values written next to each bar in Fig.~\ref{co_segregation} (a)) indicates that the most favourable segregation pair is (``0b",``0a"). 
For fcc GB, the presence of the first Zn atom at site ``0b" makes segregation of a second Zn atom unfavourable, except to the two further sites in the GB plane (sites ``0e" and ``0f"). The effect is more pronounced for the site ``0e", which was a compression site (smaller Voronoi volume than fcc bulk) before the insertion of the first Zn atom at site ``0b".
The segregation behaviour of the second Zn atom implies that the Zn-Zn interaction is slightly attractive in both bcc and fcc GBs, which is due to the gain in the elastic contribution - the first Zn in the GB expands the lattice in the vicinity of the adsorption site and weakens the neighbouring Fe-Fe bonds. This leads to an increase in the bond length (i.e., larger Voronoi volume) and hence a more favourable condition to accommodate the next Zn.    

\subsection{Grain boundary embrittlement by Zn}

\begin{figure}
\centering
\includegraphics[width=0.5\textwidth]{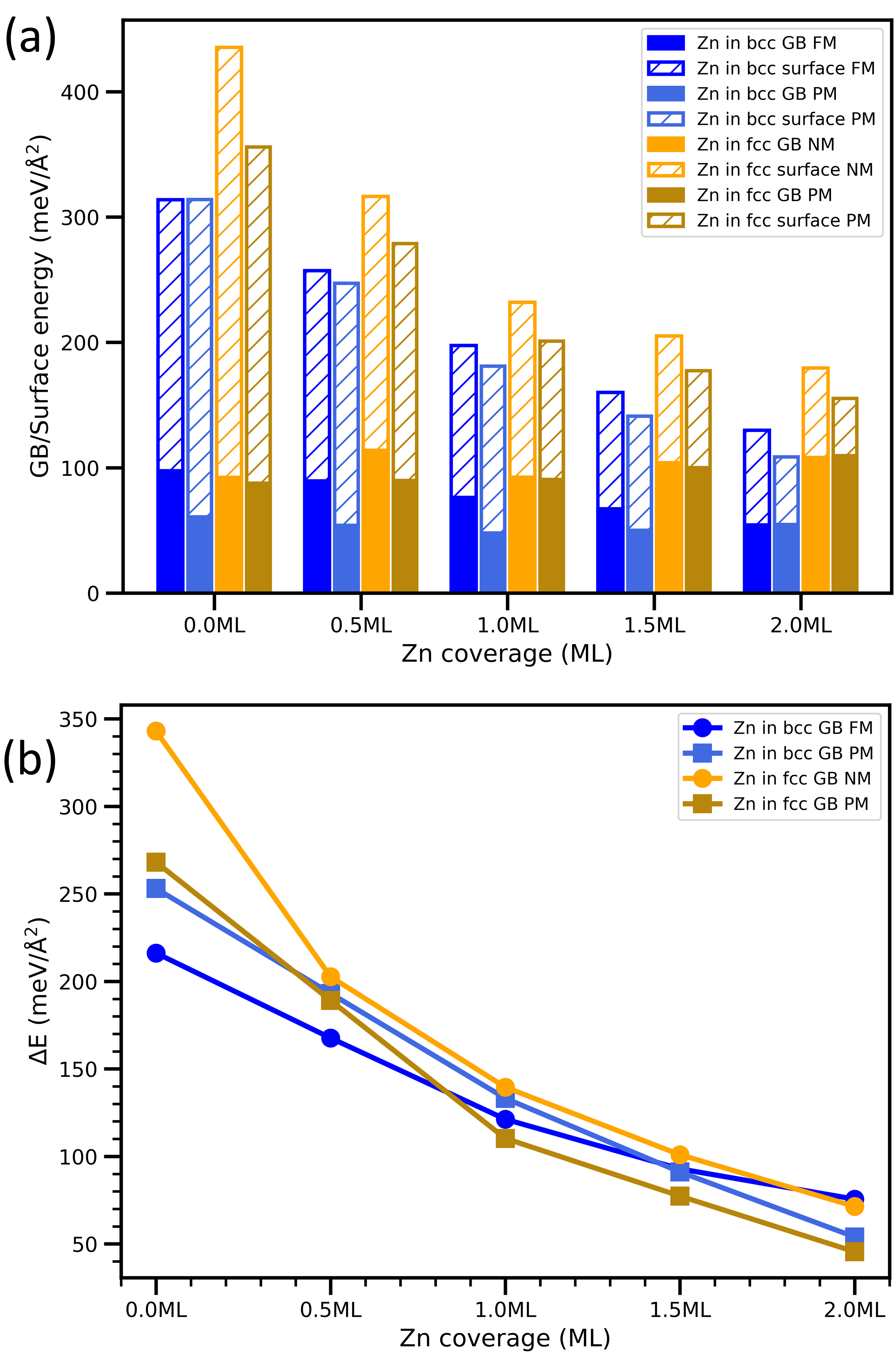}
\caption{\label{interface_energy} (a) GB energy (solid bars) and corresponding doubled surface energies (dashed bars) for bcc FM, bcc PM, fcc NM and fcc PM $\Sigma 5$ GB for different Zn coverages. (b)  Difference between the doubled surface energies and the corresponding GB energy (cf. Eq.\ref{rice_wang}) at the same Zn coverage. The energy difference is decreasing rapidly with increasing Zn coverage, clearly indicating the weakening effect Zn has on the GB.}
\end{figure}

A theoretical approach to study the wetting and embrittling effect of solutes at grain boundaries is to compare the segregation energy of the solute to both  the GBs and their corresponding surfaces. Such an approach was already used by K.~Bauer \textit{et al.} and D.~Scheiber \textit{et al.} \cite{bauer2015first,scheiber2020influence} to investigate Zn induced embrittlement behaviour in the bcc Fe $\Sigma 3$, $\Sigma 5$ and $\Sigma 9$ GBs. Here, we applied this approach to compare the Zn induced embrittlement in bcc and fcc $\Sigma 5$ GBs of Fe, but focus on the impact of magnetism. 

Accordingly, we calculate the energies for both, the pristine and Zn covered GB using Eq.~(\ref{energy_gb}). Analogously, we consider also the corresponding pristine and Zn covered surfaces and calculated the respective surface energies using Eq.~(\ref{energy_surf}). 

The way in which we evaluate these energies, will reflect the source of Zn atoms. If we assume, that the Zn atoms originate from the bulk of Fe, their amount is limited, as we are looking at a closed system. If we consider that the Zn atoms originate from the coating, we are looking at an open system and an unlimited amount of Zn atoms provided by this reservoir. We incorporate the Zn reservoir, i.e., the source, in terms of a chemical potential for Zn ($\mu_{\rm Zn}$) in Eq.~\ref{energy_gb} and Eq.~\ref{energy_surf}. Let us first assume that Zn originates from Zn in it's bulk hcp phase, which acts as the reservoir for Zn atoms. This gives an upper limit (Zn-rich conditions) for $\mu_{\rm Zn}$ in the Zn chemical potential space. 

The GB and corresponding surface energies obtained for this upper limit of the Zn chemical potential are shown in Fig.~\ref{interface_energy}(a). The energy of the bcc FM GB is decreasing with increasing Zn coverage (dark blue solid bars). At the same time the energy of the corresponding surfaces is decreasing even more rapidly with increased Zn coverage (dark blue dashed bars). The GB energy of a pristine bcc PM GB is significantly lower than of the FM GB, which decreases even further for GBs covered by 1 ML Zn and increase slightly when covered by 2 ML Zn (light blue solid bars in Fig. \ref{interface_energy}). 
The energy of the pristine bcc PM surface is the same as of the FM surface. However, with increasing Zn coverage the surface energy in the PM case is decreasing faster than in the FM case (light blue dashed bars). 
For fcc NM GB, we found a slight increase in the GB energy with higher Zn coverage (solid orange bars), while the energy of the corresponding NM fcc surfaces is decreasing with increasing Zn coverage (dashed orange bars). The GB energy of NM and PM fcc GB are similar. However, the corresponding surface energy for the PM case is always lower than for the NM case for all considered Zn coverages. Such trend in interface and surface energy as a function of Zn coverage implies stronger Zn induced embrittling effect in the considered fcc GB in comparison to the bcc GB.

To analyze the embrittlement effect at Zn-rich conditions, i.e. the upper limit of the Zn chemical potential, we plot the energy difference between each calculated GB energy and its surface energy counterpart for both bcc and fcc GBs as a function of Zn coverage for all considered magnetic configurations. As seen in Fig.~\ref{interface_energy}(b), at 0 ML coverage the energy difference between the GB and surface energy in the fcc NM and bcc FM case is much larger than for the  PM GBs. In the latter case the differences in energies is almost the same for the two crystal structures. The addition of Zn reduces the difference between interface and surface energy for all GBs, clearly reflecting the embrittling effect of Zn. Fig.~\ref{interface_energy}(b) clearly shows that increased Zn coverage has a higher impact on fcc compared to bcc GB, i.e. the difference decreases more rapidly, indicating a higher susceptibility towards Zn induced weakening for fcc GB compared to bcc GBs. Furthermore, there is a systematic offset, between the NM and PM configuration in the case of the fcc GB, while for the bcc GB the PM case yields the stronger dependence on the coverage. 

\begin{widetext}
\begin{figure}
\centering
\includegraphics[width=\textwidth]{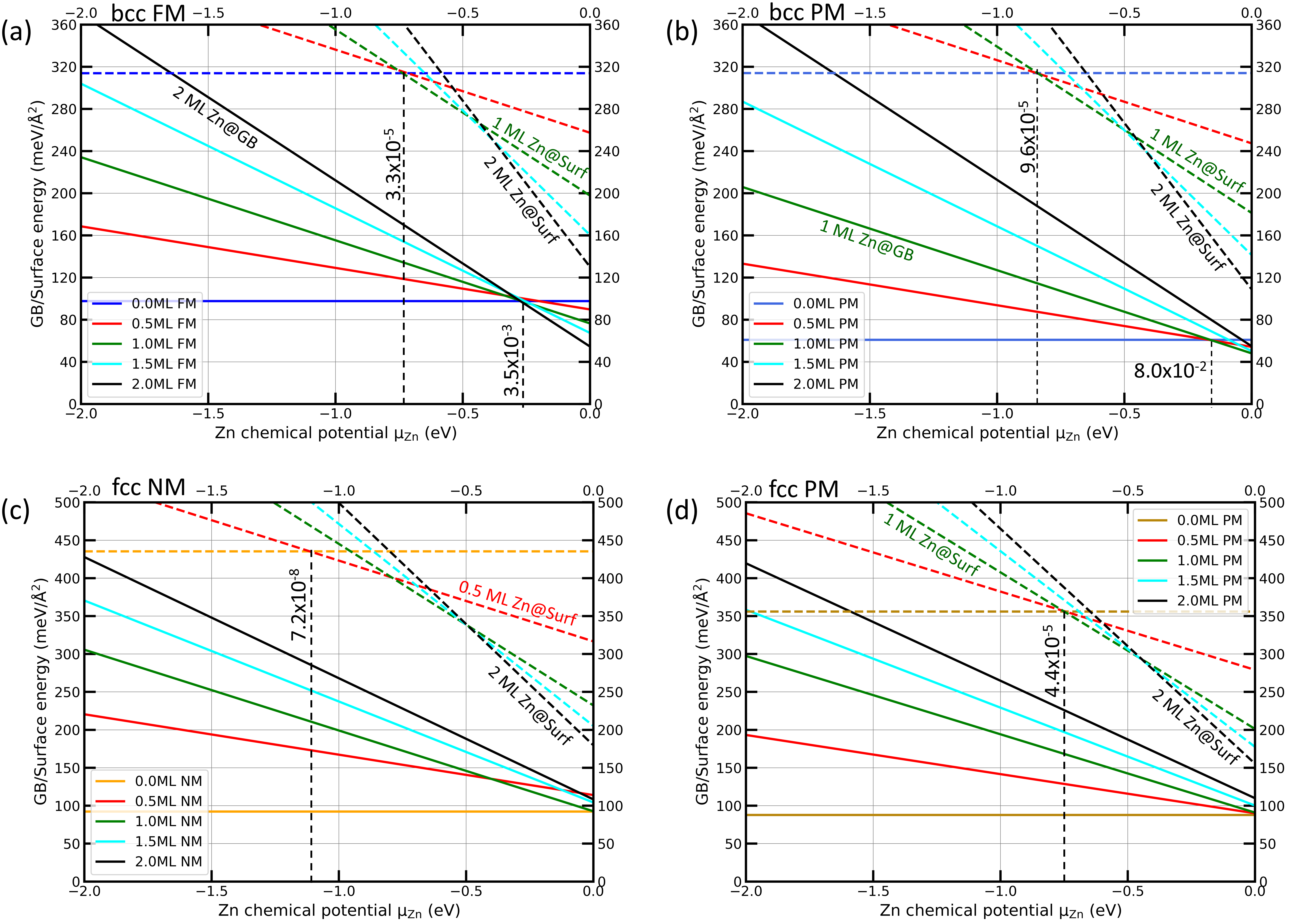}
\caption{\label{wang_rice} The Fe $\Sigma 5$ grain boundary (solid lines) and corresponding surface (dashed lines) energies as a function of the zinc chemical potential for (a) bcc FM (b) bcc PM (c) fcc NM and (d) fcc PM systems. The vertical dashed lines in each plot corresponds to the Zn concentration in ppm at which weakening of the GB starts.}
\end{figure}
\end{widetext}

Let us now extend our discussion to the consideration of other achievable chemical potential values for Zn, i.e. other conditions for the reservoir of Zn atoms. During spot welding, LME typically occurs in a temperature range from 700 to 950 $^\circ$C \cite{beal2012liquid}, which is well above the melting temperature of Zn (420 $^\circ$C). To account for these conditions, we must consider liquid Zn as being our reservoir for Zn. The DFT based value for the chemical potential of a Zn atom in a liquid Zn droplet is difficult to calculated. Another situation, where reservoir of Zn is the Zn atoms dissolved in Fe bulk will have lower chemical potential than Zn coming from bulk hcp phase. 

The Wang-Rice model as expressed in Eq.~(\ref{rice_wang}) contains a dependence on the chemical potential of Zn, using the grand canonical approximation that the system can maintain a constant Zn chemical potential by equilibrating sufficiently fast during the propagation of the crack. In Fig.~\ref{interface_energy} have used the Zn bulk phase for this purpose, which now corresponds to $\mu_{\rm Zn} = 0.0$ eV in Fig.~\ref{wang_rice}. The Figs.~\ref{wang_rice} now show the GB and corresponding surface energies as a function of the zinc chemical potential to allow a more general discussion of the Zn induced weakening of the considered bcc and fcc GBs for different Zn reservoir conditions.
Again, we plotted both FM and PM energies for bcc and NM and PM energies for fcc. 
The solid lines in these plots represents GB  energies and the dashed lines with the same colours represents corresponding surface energies.  
As evident by Fig.~\ref{wang_rice} some of the considered GB and surface defect states have for none of the chemical potentials the lowest defect energy. 
For example, the 0.5 ML Zn covered bcc FM surface does not represent a thermodynamically stable phases among the considered surface states. Similarly, 0.5 and 1 ML Zn covered bcc FM GB states are also not stable thermodynamically, but are at best metastable. Only 2 ML Zn covered bcc FM GB state is found thermodynamically stable in the considered chemical potential range (solid black line in Fig.~\ref{wang_rice} (a)). Among five considered bcc Fe PM surface states, three of them (0, 1 and 2 ML) represent thermodynamically stable defect states. Unlike the FM bcc GB, in the PM case the 2 ML bcc GB state is not stable, only 1 ML bcc PM defect state is found stable thermodynamically (solid green line in Fig.~\ref{wang_rice} (b)). 

As shown in Fig.~\ref{wang_rice} (c), all considered Zn covered fcc NM surface states are found stable except one (1.5 ML). Similarly, the considered fcc PM surface states (0, 1 and 2 ML) are also found thermodynamically stable. None of the considered Zn covered fcc NM and PM GB states represent stable thermodynamic phase. 

We can now compare the effect of Zn between the bcc and fcc crystal structures. It is evident from Fig.~\ref{wang_rice} that a substantial reduction of the bcc GB energy by Zn is observed for Zn rich conditions. Such an effect is not observed for the fcc GB within the considered range of Zn chemical potential. 
The situation is opposite for the surface states, where the reduction of the surface energy sets in at a slightly lower chemical potential for fcc as compared to bcc. 
In both cases, however, the impact of Zn on the surface state is most relevant for the onset of LME at low Zn chemical potentials. 
We calculate the critical Zn concentration at $T$ = 900 $^\circ$C, which is a relevant temperature in the context of Zn induced LME in high strength steels. Here we assumed that our system is in thermodynamic equilibrium and used $c_{\rm Zn}(\mu_{\rm Zn}) = \exp(-(E_{form}-\mu_{\rm Zn})/k_{\rm B}T)$ to calculate the Zn concentration as a function of Zn chemical potential. It appears that the onset of the bcc FM (PM) GB weakening starts at a Zn concentration of 0.3 ppm (1.0 ppm). For the fcc NM (PM), onset of GB weakening occurs at an even much lower Zn concentration of 0.07 ppm (0.4 ppm). Notably, for both crystal structures, onset of GB weakening occurs at higher Zn concentrations in corresponding PM GBs. The surface defect states always become stable at much lower Zn concentration than the GB defect states. None of the considered fcc NM and PM GB defect states are found stable over the corresponding pristine GB states in the considered chemical potential range.

\section{Conclusion}
In conclusion, we use DFT calculations and investigate the difference in behaviour of Zn at bcc and fcc GBs to understand the impact of crystal structure towards LME. The impact of magnetism is also addressed by considering different magnetic ordering of the considered GBs (FM and PM for bcc; NM and PM for fcc). To overcome the limitation of DFT based paramagnetic calculations we use the spin space averaging relaxation approach for PM GBs. Considering $\Sigma 5[100]36.8^{\circ}$ as the representative GB, we investigate the segregation behaviour of Zn for these different structural and magnetic orderings. Overall, Zn has a higher tendency of segregation to the bcc compared to the fcc GB and primary requirement for segregation is the elastic contribution for both crystal structures. Changing the magnetic phase to PM (from FM in bcc and from NM in fcc) reduces the tendency of Zn segregation to GBs.

Combining DFT calculations with a grand canonical Rice-Wang thermodynamic model, we evaluate the difference between GB and corresponding surface energies to investigate the Zn induced decohesion effect. For all considered structural and magnetic ordering, rapid decrease in surface defect states energies is the driving force for Zn induced weakening of the GBs. The critical concentration for onset of GB weakening is higher for the PM GBs compared to FM GB in bcc and NM GB in fcc. Surface defect states stabilize at lower Zn concentrations than the GB defect states, and none of the examined fcc GB defect states are found stable compared to their pristine counterparts.

\section{Acknowledgement}
We acknowledge financial support by DFG transfer project T7 (part of SFB761 ``Stahl - ab initio''). We are grateful to Dr. Osamu Waseda (Max-Planck-Institut f{\"u}r Eisenforschung GmbH, Germany), Dr.~-Ing. Konstantin Molodov (Salzgitter Mannesmann Forschung GmbH, Germany) and Dr.~-Ing. Stefanie Sandl{\"o}bes-Haut (RWTH Aachen, Germany) for fruitful discussions. We also acknowledge the scientific discussions with our colleagues in the weekly departmental meeting ``Ab initio description of iron and steel (ADIS)" and technical support from the pyiron developers.


\bibliography{references}

\end{document}